\documentclass[twocolumn,showpacs,preprintnumbers,amsmath,amssymb]{revtex4-1}
\pdfoutput=1
\usepackage{graphicx}
\usepackage{dcolumn}
\usepackage{bm}
\usepackage{hyperref}
\begin{document}
\preprint{CB001.1}
\title{General linear matrix model, Minkowski spacetime and the Standard Model}
\author{Chris Belyea}
\email{cibelyea@gmail.com}
\date{\today}
\begin{abstract} 
The Hermitian matrix model with general linear symmetry is argued to decouple into a finite unitary matrix model that contains metastable multidimensional lattice configurations and a fermion determinant. The simplest metastable state groups a two component spinor from the matrix and locally describes a Hermitian Weyl kinetic operator of either handedness on a hypercubic 3+1 dimensional lattice with general nonlocal interactions. The Hermiticity produces 16 effective Weyl fermions at long distances by species doubling, 8 left- and 8 right-handed.  These are identified with a Standard Model generation.  Only local non-anomalous gauge fields within the soup of general fluctuations can produce effective massless propagators and survive at long distances, and the degrees of freedom to support  non-anomalous gauge field subgroups of $U(8)_L X U(8)_R$ are present.  Standard Model gauge symmetries associate with particular species symmetries, for example change of QCD color associates with permutation of doubling status amongst space directions.  Vierbein gravity is probably also generated. Low fermion current masses can arise from chiral symmetry breaking solutions of the fermion self-energy Schwinger-Dyson equations, generating W mass and composite Higgs states, similar to a scheme proposed by Gribov. Specific higher dimensional lattices with larger spinors are  potentially stable but produce non-Riemannian spaces without conserved quadratic distances. However if the extra dimensions are compactified, the Minowski space persists at low energy accompanied by SM generations, potentially doubled further by duplicate zero modes in the compact directions to generate dark matter.  The model is conjectured to have an origin in infinite dimensional conformal invariance and the concept of Bare Particulars.

\end{abstract}
\pacs{12.90.+b, 11.10.-z, 11.10.Lm, 11.10.Kk, 11.15.Ha, 11.25.Hf, 11.25.Mj, 11.25.Yb, 12.10.-g, 12.10.Dm}

\maketitle
\tableofcontents
\section{\label{sec:1}Introduction}

If there is a fundamental physical theory it is expected to be based on a simple and metaphysically compelling principle, while at the same time providing an account of the numerous very specific features we observe in low energy physical laws.  By anthropic selection, some but hopefully not all of these features could be accounted for contingently as those of a habitable stable configuration in a landscape of other  stable configurations, some habitable and some barren. According to string theories that landscape is rich and highly varied, with our 3+1 dimensional configuration of interacting Standard Model fermion generations not obviously typical or naturally selected from amongst almost countless alternative varieties.    

Such a scenario is not strictly objectionable scientifically but there has been criticism on the grounds that too much anthropic freedom would render the fundamental theory unfalsifiable. There may be an alternative model to string theories that is simpler and relies less on anthropic selection, more directly pointing to our universe amongst a set of not very dissimilar alternative configurations.  The purpose of this paper is to argue that the simplest most symmetric of all Hermitian matrix models, which may have an origin in infinite dimensional conformal invariance and the metaphysical concept of Bare Particulars, could provide this alternative.

Hermitian matrices that represent dynamic connection between all points in a discrete space are a promising device to model the emergence of locally connected spaces.   While not providing the sought for unification in string theory, they have been used to model the dynamical emergence of a world sheet in $D \leq 2$, and in a number of ways to model target space dimensions\cite{Seiberg2006}. Independently of string theories, matrix models have been recently used to study spontaneous phase transitions into local multidimensional spaces\cite{Delga2008,Delga2009}.

In this paper the matrix rather than the world sheet is followed as the key element of interest in building a fundamental theory and, in keeping with the expectation of simplicity, the object of study is the unique Hermitian matrix model having general linear symmetry. Its scale invariant pathology decouples leaving a well-defined submatrix system with unitary symmetry operating on half the number of points, dominated by first order difference operators. When the total number of points N is odd the submatrix system has a determinant that represents strongly coupled nonlocal fermions. Statistical weight arguments suggest that only certain multidimensional configurations potentially are metastable spacetimes with quadratic distance measures.  The dimensionality, signature, particle spectrum, and effective interactions are tightly constrained and all look very like our Universe.

\section{\label{sec:2} The General Linear Matrix Model}
\subsection{\label{sec:2.1}Construction}

 If $\Lambda$ is an N by N general complex matrix and $H$ is Hermitian, then the transformation 
\begin{equation} 
H^\prime = \Lambda^\dagger H \Lambda 
\end{equation}
preserves Hermiticity.  Physics is proposed to be represented by a Hermitian matrix model invariant under (1) for invertible $\Lambda$, with N  either infinite or large enough to provide all the points needed for a discrete Planck scale spacetime. 

The flat integration measure in terms of real and imaginary components is

\begin{equation}
DH = \prod_{i} dH_{iiR} \prod_{i<j} dH_{ijR}dH_{ijI}.
\end{equation}

Under (1) this transforms as $DH^\prime = \det (\Lambda^\dagger\Lambda)^N DH$. This can be seen by considering that the general complex matrix $M=H+iA$ where A is also Hermitian has Jacobean $J^2$ where $J$ is the Jacobean for $H^\prime$ or $A^\prime$. The Jacobean for $M^\prime$ is easily calculated by multiplying $M$ first on the right by $\Lambda$ and then on the left by $\Lambda^\dagger$ contributing $\det(\Lambda^\dagger \Lambda)^{N}$ at each stage. The invariant measure $\hat{D}H$ is therefore $DH/\det{H^N}$.

The unique and seemingly barren partition function is formed out of just the invariant measure which we call $H$-theory

\begin{equation}
Z=\int \hat{D}H=  \int \frac{DH}{\det H^N}
\end{equation}

\subsection{Finite effective subsystem}

We take the standard path of examining the eigenvalues of the configurations that dominate the partition function, and since there is a unitary symmetry only the eigenvalues control the weight. 

Using the techniques of random matrix theory\cite{Dyson1962,Forrester2010,Zee2003,Mehta2004} $H$ is diagonalized to extract the eigenvalues to obtain 

\begin{equation}
Z_{H}= V \int D \lambda s(\lambda)^N e^{-V(\lambda)} 
\end{equation}
 where $V=\int DU$ is the constant volume of the unitary group,$s(\lambda)$ is the product of the signs of the eigenvalues and

\begin{equation}
V(\lambda)= \frac{N}{2}\sum^N_{i=1} \log \lambda_i^2  -  \sum^N_{i<j=1} \log (\lambda_i - \lambda_j )^2
\end{equation}
 
The value of the eigenvalue integral, having a scale invariance, is not well defined, but nonetheless we can show that a well defined system  consistently decouples. 

The commonest or highest weight eigenvalue configurations can be found by minimizing $V$. $V$ is analogous to the potential energy (no kinetic energy) in a classical one dimensional Dyson gas of $N$ mutually repelling point particles of  position coordinates $\lambda_n$ and charge $+1$  attracted by a fixed nucleus at $\lambda=0$ of charge $-N/2$. The mutual repulsion coming from the second term in $V$ comes from  the measure $DH$ due to the Van der Monde determinant Jacobean.  This is universal in Hermitian matrix models and normally leads to constant low-lying eigenvalue density for all $V$ that are well-behaved at zero, unlike this case. 

The singular behaviour at zero can be shown to decouple by using the gas analogy. There is a balance between the attractive power of the pole and the repulsive power of the random statistics, so that only half the eigenvalues collapse.  If $N$ is even, $N/2$ eigenvalues will collapse into the nucleus to form an infinitessimal neutral atom and if $N$ is odd $(N+1)/2$ eigenvalues will collapse to form an infinitessimal ion of charge $+\frac{1}{2}$.  The remaining eigenvalues will interact among one another as a free gas either without effect from the neutralized atom (N even) or under the milder repulsive influence of the central $+1/2$ ionic charge (N odd).   

In either case the particular infinitessimal values of the bound eigenvalues which collapse to zero have no effect on the free eigenvalues, the only effect being the central charge neutralization. The free eigenvalues can only see the aggregate central charge.  The divergent dynamics of the central atom therefore decouples and the free eigenvalues are  part of an independent submatrix system. If N is odd, where there is a residual central repulsive ion, the remaining $N^\prime=(N-1)/2$ eigenvalues correspond to the system $\int DH \det H$, the residual determinant being suggestive of a pregeometric system of strongly coupled fermions.  If N is even, where the central atom is neutral and invisible, the remaining $N^\prime=N/2$ eigenvalues correspond to the system $\int DH$ which unlike the former case is empty of fermionic matter. 

Before proceeding we need to go back and make a scale transformation in the eigenvalue integral. The system $\int DH \det H$  is dominated by the free gas of eigenvalues that now fly off to infinity under unconstrained mutual repulsion. However this infinity can also be decoupled by scaling it down into the central nucleus infinity by noting that the eigenvalue integral is of the special form 

\begin{equation}
 \lim_{p\rightarrow\infty} \int^p_{-p} D\lambda F(\lambda)
\end{equation}
where $D\lambda F(\lambda)$ is scale invariant. Rescaling $\lambda_i\rightarrow\lambda_i s/p$ with $s$ finite leads to  
\begin{equation}\label{scaleeq}
 \lim_{p\rightarrow\infty} \int^p_{-p} D\lambda F(\lambda) = \int^s_{-s} D\lambda F(\lambda)
\end{equation}

Essentially the integration limit is arbitrary due to the scale invariance. Choosing $s$=1  the previous analysis is repeated. The well-behaved subsystem for odd $N^\prime$ has Dyson gas potential within the box $-1\leq \lambda_i < 1$ of 

\begin{equation}
V(\lambda)= -\frac{1}{2}\sum^{N^\prime}_{i=1} \log \lambda_i^2  -  \sum^{N^\prime}_{i<j=1} \log (\lambda_i - \lambda_j )^2
\end{equation}

which, since the function $\lambda^{2B}$ converges to the box function for large integer B, is generated by the reduced matrix model as $B\rightarrow\infty$ 

\begin{equation}
Z_R=\int DH \det H e^{-trH^{2B}}.
\end{equation}

The regulator reduces the symmetry from unimodular to unitary.

\subsection{Dominant Eigenvalue density}

The residual $+1/2$ central charge term in $V$ from the determinant prevents zero eigenvalues and affects the first few eigenvalues neighbouring zero, but at large $N^\prime$ this term has negligible effect on the overall shape of the eigenvalue density, which is swamped except for these near-zero eigenvalues (representing cosmic distances if distance is inverse eigenvalue) by the mutual repulsions amongst the gas of $N^\prime$ charge +1 particles. Consequently we can neglect this term in determining the dominant eigenvalue density, which is therefore controlled solely by the measure, i.e. the statistics of random configurations, and the confining box. 

It is argued later that the determinant creates the fermions of the Standard Model and their strong non-local interaction condenses to local propagating gauge and gravitational fields. The negligible local effect of this determinant at the pregeometrical stage shows that the statistical influence of the matrix integral that creates the local spacetime background in this model is a far stronger force than the forces we know about.

If the eigenvalues are labelled $\lambda(x)$ in the continuum limit with $x$ a real number ranging lowest to highest from $0$ to $1$, the dominant density of eigenvalues $\rho(\lambda)=dx/d\lambda$ can be determined. The force $-\partial V/\partial\lambda_i$ on each eigenvalue $i$ due to the others must balance, and so the eigenvalue density near the box boundary must go to infinity due to the force from one direction. We guess the simple form $\rho(\lambda)\propto(1-\lambda^2)^{-1/2}$ in analogy with the Wigner semicircle law  $(a^2-\lambda^2)^{1/2}$ for the Gaussian confining potential. This is confirmed by  writing the continuum version of the force balance equation as

\begin{equation}
    P \int^1_0 dx \frac{1}{\lambda(x)-y} = 0   
\end{equation}

\begin{figure}
\includegraphics{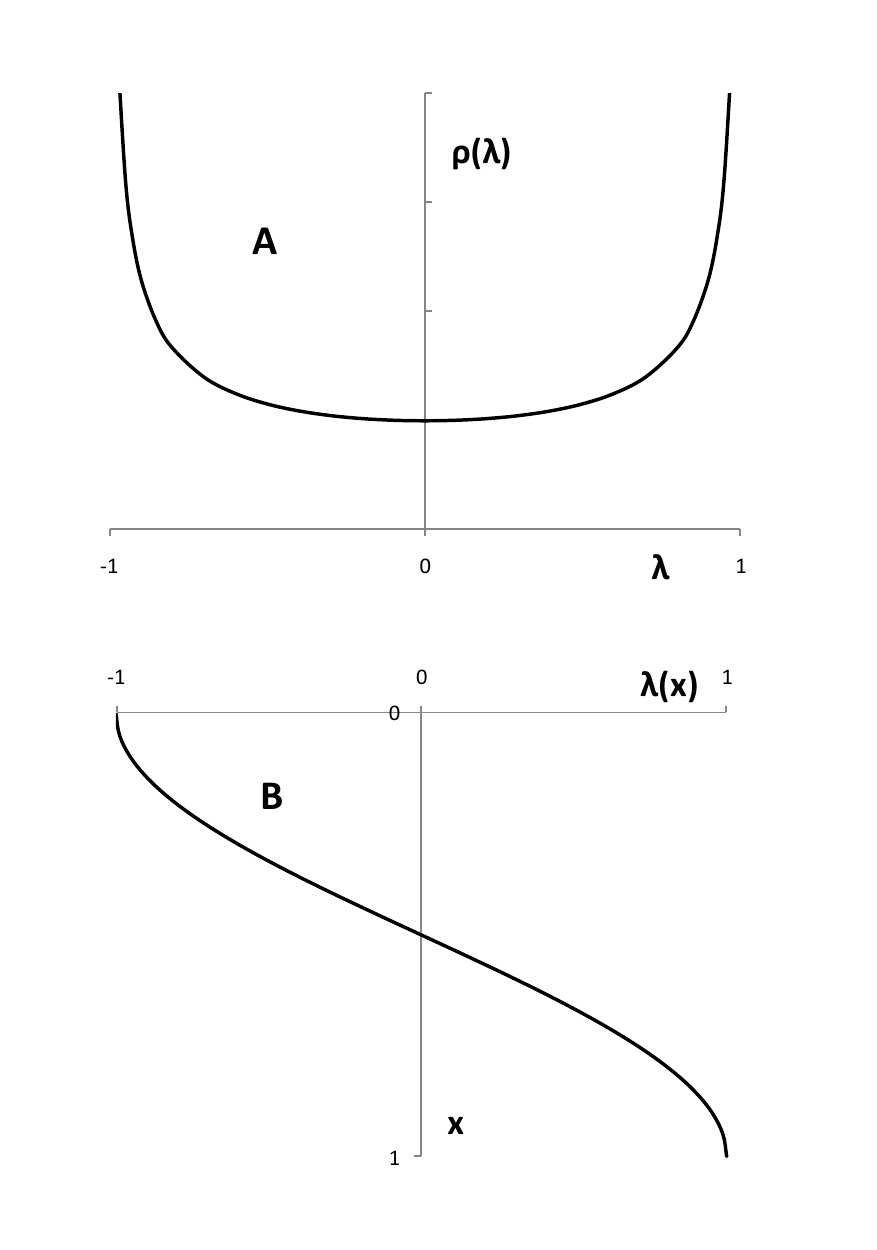}
\caption{\label{fig:domeig} Dominant eigenvalues for the box potential. A: eigenvalue density ; B:  eigenvalues ordered lowest to highest}
\end{figure}

$\forall y\in(0,1) $ where $P$ is the Cauchy Principal value, then showing by contour integration that $\lambda=-\cos(\pi x)$ is the solution. See Figure \ref{fig:domeig}.

\subsection{Local Difference Operator Form}

\begin{figure}
\includegraphics{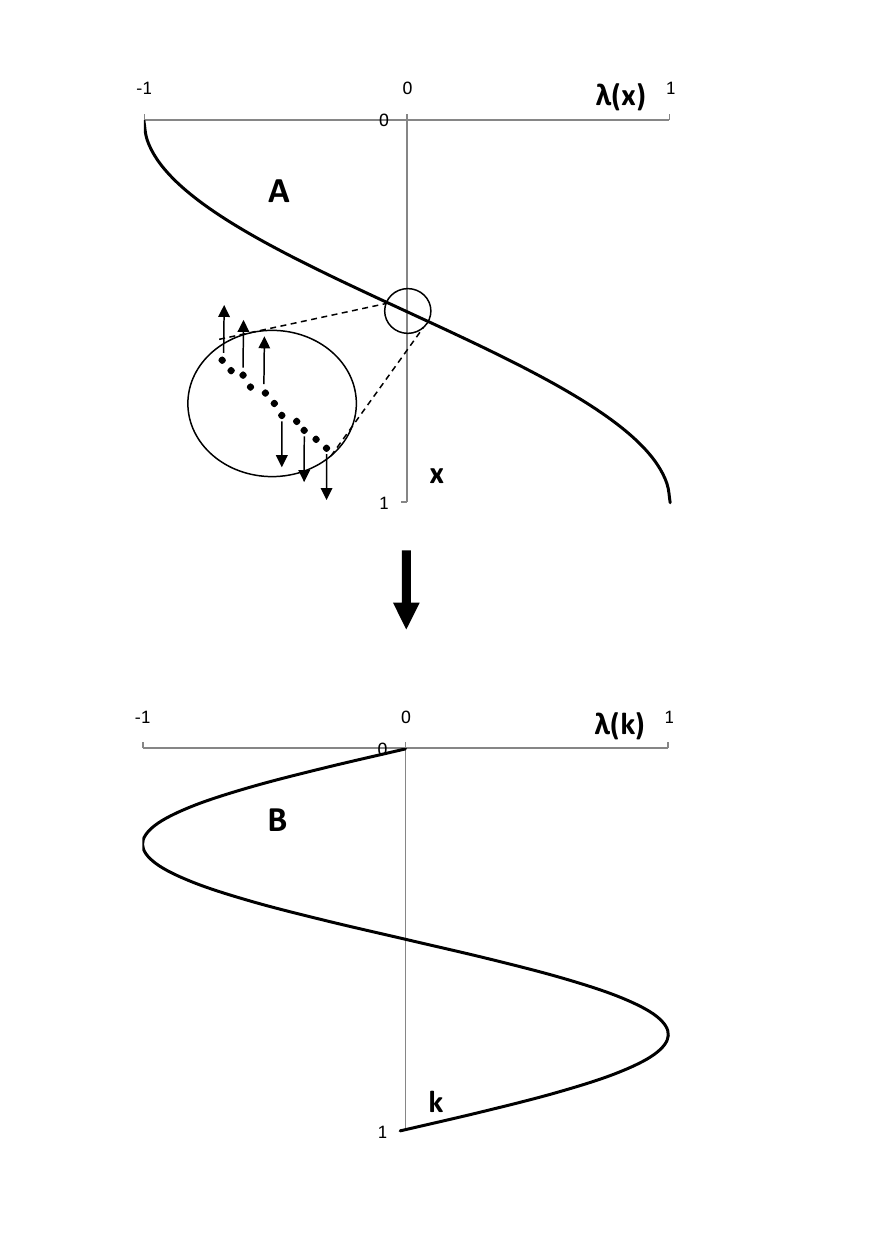}
\caption{\label{fig:reorder} Unitary transformation to local difference operator - A: lowest to highest order showing every second eigenvalue re-ordered B:   sinusoidal result}
\end{figure}

The constant eigenvalue density near zero is a universal feature of Hermitian matrix measures and causes the dominant configurations, in the absence of special or tuned potentials, to be unitarily equivalent to a local first order difference operator.  If H is expanded about its background value as $H=H_B+H_I$ then by the unitary symmetry $H\rightarrow UHU^\dagger$,

\begin{equation}
 \det (H_I+H_B) = \det(H_I+UH_BU^\dagger)
\end{equation}

and so all unitarily equivalent $H_B$ describe the same physics. The local difference operator will be always be the one from which the physical law is written, so we perform the unitary transformation of re-ordering the eigenvalues away from their lowest to highest order to form a momentum space operator that has the period of a full sinusoidal cycle as shown in Figure \ref{fig:reorder}, which for smoothly spaced eigenvalues corresponds to the Hermitian nearest-neighbour first order derivative  $i\Delta_{xy}=\frac{i}{2}(\delta_{y,x+1}-\delta_{x,y+1})$. Interestingly in this case there are no further modes because the shape of the momentum space eigenvalue function is precisely sinusoidal. This is different from other matrix models, which like the Gaussian case have decreasing components of third-, fifth- and higher odd nearest-neighbour terms, due to the difference of their eigenvalue density from the form that gives pure sinusoidal shape.

Throughout we ignore overall normalisation constants in $H_B$ since the constraint on the eigenvalues of  $H_I+H_B$ is unambiguous.

The configuration that minimises the Van der Monde term will have smoothly spaced eigenvalues and be precisely the first order difference operator on a one dimensional nearest-neighbour lattice. 

Typical configurations of H when $N$ is large  (i.e. common configurations chosen at random from H) will still be approximately the nearest neighbour first order difference operator but with random weight between adjacent points that reflects a random spacing $s$ between adjacent eigenvalues roughly according to the Wigner surmise\cite{Mehta2004} probability distribution $se^{-s^2}$. 

Neither the minimizing nor typical configurations of $H$ are a multidimensional spacetime, but this is irrelevant to whether physics is described by the general linear model. We find ourselves in a highly untypical place on the surface of Earth because the vastly more typical places in outer space are inhospitable, and by a weak version of the anthropic principle the same can apply to the configuration of a habitable potentially stable multidimensional space-time in $H$ which is otherwise dominated by the uninhabitable one-dimensional class. Since the matrix connects all points arbitrarily, any Hermitian multidimensional matrices are configurations within the system. Our Universe with its low energy density after the Big Bang can be represented by a decoherent region around a multidimensional configuration as long as the configuration is stable against small local fluctuations, and the local fluctuations are all that survive at low energy. 

\section{Spinor Spacetime Lattices}

\subsection{Metastability Criteria}

A definitive demonstration of metastability of any particular multidimensional configuration is not within grasp, but criteria based on statistical weight can be given. 

Since the first order difference operator lattice is stabilized by the mutual repulsion of eigenvalues in the measure $DH$ for a large Hermitian matrix, the same statistical forces apply to any large Hermitian submatrix of $H$.  Consequently in any metastable configuration, every point in the vector space will have local neighbour connections as part of one or more locally stable first order difference operators. The larger the identifiable submatrix, the more stable and the greater statistical penalty to be paid by disrupting the point from its neighbour connnections. 

Consequently a spacetime with simple regular connectedness such as a hypercubic lattice threaded at every point with first order difference operators inherited from the one dimensional configuration is favoured over random connectedness that will strand many points from participation in any local difference operator. For a D dimensional spacetime, each point on the lattice is passed through in D directions by D individually stable first order difference operators and will participate independently in crossing local submatrices.  For this to occur, the D operators must be independent at each spacetime point which requires that a number of points from the vector space are gathered together to form an internal spinor degree of freedom. 

While a submatrix might not necessarily have its eigenvalues bound by the box potential, there will be an effective confinement provided by the environment of the matrix and its overall box potential bound. Consequently there may be departure from the pure nearest neighbour difference operator of the one-dimensional configurations and inclusion of third, fifth and greater neighbours, but nonetheless a local difference operator is expected.  We also expect an element of variation in the strength of each link, in particular since the eigenvalues of an exactly regular multidimensional lattice are have degenerate points associated with permutation symmetries, forbidden by the eigenvalue repulsion.

One major consequence of the submatrix argument is that large zero Hermitian submatrices are highly unstable.  This exclusion criterion powerfully restricts the potentially stable configurations.

\subsection{4-dimensional Minkowski Weyl lattice}

In what follows spinor spacetimes will be considered with total numbers of configuration space points $N^\prime$ even, which is necessary since spinors have two points per spacetime point. A single odd spacetime point, which must be part of the total $N^\prime$ system if $N^\prime$ is odd, can always be added at a boundary.

We begin with the simple Hermitian hypercubic lattice representation of the left-handed Minkowski Weyl kinetic operator in four dimensions $\bar{\sigma}^\mu i\Delta_\mu$, which is clearly a configuration in $H$ since any $H$ can be decomposed into four Hermitian matrices by $H=\bar{\sigma}^\mu H_\mu$.  

Assuming each dimension has L points, each matrix $\Delta_\mu$ contains $L^3$  distinct one dimensional difference operators parallel to the $\mu$ axis, one for each for each coordinate value of the other three dimensions, e.g 

\begin{equation}
\Delta_{0xy}= \delta_{x_1y_1}\delta_{x_2y_2}\delta_{x_3y_3}\Delta_{x_0y_0}
\end{equation}

\begin{figure}
\includegraphics{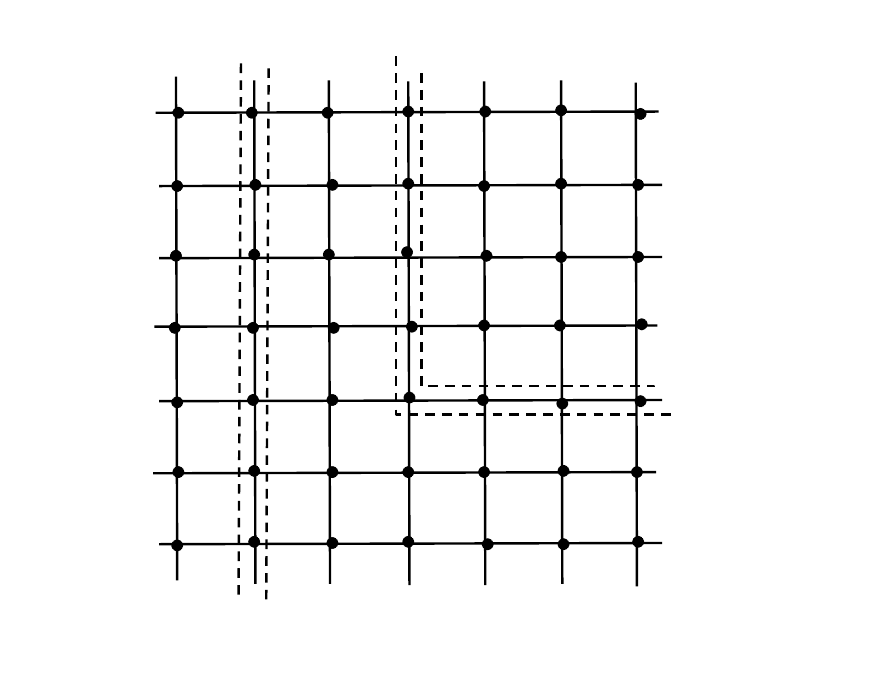}
\caption{\label{fig:lat1} submatrices that contribute to stability}
\end{figure}

There is no spacetime point that is not connected into such a difference operator. Any removal of a neighbour link will incur a large statistical penalty in one of the large stable one dimensional difference operators extending along one of the axes. As shown in Fig. \ref{fig:lat1} the submatrices that influence the stability of the lattice are not just straight sets. There are submatrices linking a set of points that turn one or more corners, where it can be shown (at least for the nearest neighbour case) by transformations on the spinor matrices that the eigenvalues associated with such crooked sets are the same as the straight sets.  These crooked sets help to favour the simultaneous diagonalisation of all the $H_\mu$ that makes all dimensions simultaneosly local (e.g. $\left[\partial_\mu,\partial_\nu\right]=0$) and also to favour the same link length in each dimension. 

From this example it follows that two or three spacetime dimensions is highly unstable, since we still need a spinor index and the spinor matrix will be unsaturated, with at least one of the $H_\mu$ or a linear combination equal to zero. This does not however rule out toroidal compactification of the four dimensional case. Similarly other unsaturated spinor matrices such as $\sigma^\mu\oplus\bar{\sigma}^\mu$ necessary for a fundamental Dirac particle or any other unsaturated multiplicity are also unstable. 

Euclidean space Weyl operators are also not allowed because they are non-Hermitian. 

In summary, the Minkowski right or left-handed 3+1 dimensional Weyl class is the {\it only} potentially stable fermion operator in four dimensions, and three and two dimensions are unstable. 

Nielsen and Rugh\cite{Nielsen1992,Nielsen1994} have previously made related general observations about the special dynamical stability of the 0+1 and 3+1 dimensional Weyl operator. 

The way in which multidimensional spacetime is treated here, as a metastable lattice configuration of threaded first order difference operators stabilized by the quantum measure $DH$, contrasts with other various matrix models previously applied to string theory that have been specifically designed with a ``time'' coordinate $M(t)$ and a quantum mechanical action to generate an extra dimension from the matrix eigenvalues\cite{Martinec2004,Banks1999} or that generate a fuzzy sphere using a partition function over multiple matrices weighted by an SO(3) invariant action\cite{Delga2008,Delga2009}.  

The key assertion is that there is enough localising power in the measure alone to build a metastable multidimensional lattice, which is the simplest potentially metastable configuration to potentially contain life and happens to be Minkowski space. The analysis that follows argues that this configuration might actually contain the Standard Model and gravity.

\subsection{16 Fermions from Species Doubling}

For a general background configuration $H_B$ the determinant can be converted to an action with Grassmann spinor fields $\psi_{\alpha x}$ and $\bar{\psi}_{\alpha x}$ and the reduced system expanded about $H_B$ up to a multiplicative constant is

\begin{equation}
Z_R=\int_B DH_I \int D\psi D\psi e^{i\bar{\psi}(H_B+H_I)\psi}
\end{equation}

where the $B$ label on the integral records the regulator that confines the eigenvalues of $H_B+H_I$, and $i$ is in the exponent to reflect the conventional format, discarding the consequent $i^{-N}$ multiplier out the front of $Z$. Doing a perturbation expansion about $H_I=0$ when $H_B$ is the Minkowski lattice Weyl operator produces the lattice Weyl equation $i\bar{\sigma}^\mu \Delta_\mu \psi =0$ from the stationary point.

Because the Minkowski Weyl operator on the lattice is local, first order and Hermitian\cite{Nielsen1981}, it produces species doubling so that at the stationary point in the long distance regime there are actually 8 left handed and 8 right handed species. In momentum space it is the zero values of $\sin p_\mu$ at $p_\mu=\pi k_\mu = \pm \pi$ as well as at $p_\mu=0$  that are necessary to provide the local derivative (see Fig \ref{fig:reorder}B above) that result in $2^4$ states. Doubling has previously been seen as a problem for lattice simulations that add unwanted degrees of freedom, but in this specific case, where there are no bare gauge propagators, the extra degrees of freedom can be the various quarks and leptons of a generation.

Species doubling is not affected by adding 3rd, 5th or higher order modes in each dimension, as they just change the shape away from pure sinusoidal while leaving the zeroes unchanged, as in the Gaussian matrix model, nor is it affected by keeping the hypercubic connectedness but adding randomness onto the link weights\cite{Kieu1994}. It has also been noted in the fermionic sector of string matrix models\cite{Sochichiu2000}.

The extra species can be absent in lattices where the points are randomly connected\cite{Kieu1993,Kieu1994}, and there are also specially designed variations to the hypercubic lattice derivative that can eliminate the doubling species\cite{Luscher1998}, useful in computations in the lattice approximation to QCD. Neither of these apply to this necessarily spinor-dense hypercubic lattice.

The species are best expressed for the current purposes in position space. We write the $H_I=0$ action as

\begin{equation}
S_0= \sum_{xy} \bar{\psi}_x i \bar{\sigma}^\mu \frac{1}{2}(\delta_{x,y+\hat{\mu}}- \delta_{y,x+\hat{\mu}})\psi_y.
\end{equation}

where $\hat{\mu}$ is a unit lattice displacement in the $+\mu$ direction.

The extra species in position space are rapidly varying in one or more direction - e.g. $(-1)^{y_2}\chi_y$ solves the stationarity criterion if $\chi_y$ is a solution of 

\begin{equation}
\sum_y i\bar{\sigma}^\mu (-1)^{\delta_{\mu,2}}\frac{1}{2}(\delta_{x,y+\hat{\mu}}- \delta_{y,x+\hat{\mu}})\chi_y = 0
\end{equation}

which means that $\chi^\prime=\sigma^2\chi$ obeys the right handed Weyl equation $i\sigma^\mu \Delta_\mu \chi^\prime =0$. Consequently the low energy/momentum region contains not just slowly varying $\psi_y$ but also rapidly varying regions such as $(-1)^{y_2}\chi_y$, where $\chi_y$ is slowly varying.

We can express the fermion field in terms of slowly varying species components $\psi^d_x$ and $\bar{\psi}^d_x$ where $d=(d_0,d_1,d_2,d_3)$ by

\begin{equation}
\psi_x=\sum_{d}  (-1)^{d \cdot x}f_d\psi^d_x\ \ ; \ \ \ \ \bar{\psi}_x=\sum_{d}  \bar{\psi}^d_x \bar{f_d}  (-1)^{d \cdot x}
\end{equation}

where each $d_r$ takes the value 0 or 1, $d \cdot x = \sum_r d_r x_r$ and $f_d$ and $\bar{f}_d$ are spinor matrix factors to be determined. The slowly varying components need to be defined in terms of general functions of x convoluted with a smoothing function that confines the momenta to the region $\left[-\pi/2,\pi/2\right]$, but there is no need for this detail here.  Referring to the action it is easily seen using identities such as

\begin{equation}
(-1)^{d \cdot x} \delta_{x,y+\hat{\mu}} (-1)^{{d^\prime} \cdot y}  = 
 (-1)^{(d+d^\prime)\cdot x} (-1)^{\sum_s d^\prime_s\delta_{\mu,s}} \delta_{x,y+\hat{\mu}} 
\end{equation}

that the oscillatory component vanishes only when $d=d^\prime$ and for slowly varying functions the cross terms linking $\bar{\psi}^d$ with $\psi^{d ^\prime}$ for $d\neq d^\prime$ are therefore cancelled to zero and the continuum $H_I=0$ action is 

\begin{equation}
S_0= \int d^4x \sum_d \bar{\psi}^d \bar{f}_di \bar{\sigma}^\mu (-1)^{d_s\delta_{\mu,s}} \partial_\mu f_d \psi^d.
\end{equation}

Choosing $f_d=i^{d_0}(\sigma^1)^{d_1}(\sigma^2)^{d_2}(\sigma^3)^{d_3}$ and $\bar{f}_d=(\sigma^3)^{d_3}(\sigma^2)^{d_2}(\sigma^1)^{d_1}i^{d_0}$ casts the spinor matrices into the standard form. For example, in the above example $\chi= \psi^{0010}$ and $f_d=\bar{f_d}=\sigma^2$. Note that in general $\bar{f}_d \neq f_d^\dagger$ due to the $i^{d_0}$ terms. The final canonical continuum form is  

\begin{equation}
S_0= \int d^4x \sum_{d_R} \bar{\psi}^{d_R} i \sigma^\mu  \partial_\mu \psi^{d_R} + \sum_{d_L} \bar{\psi}^{d_L} i \bar{\sigma}^\mu  \partial_\mu \psi^{d_L}
\end{equation}

where the $d_L$ have an even number of ones and the $d_R$ an odd number.

\subsection{Interspecies Interactions}

Interactions between doubling species have been rarely considered before and only in the context of a lattice defined with a plaquette action, i.e. a bare gauge kinetic term, in which case the species interactions are off-shell artifacts.  In this case, there are no bare gauge kinetic terms and the interactions actually create the mass shell of the effective gauge propagators.  Rapidly varying $H_I$ fluctuations produce couplings between the species at low energy if the rapid variation in $H_I$ cancels the rapid variations from the cross-terms in the species, which will be shown explicitly in the next subsection. 

In general there are a huge number of interactions in $H_I$, both local and nonlocal, most of which are neither Lorentz nor gauge invariant. A fluctuation can only become an observable low energy field   if there is  a long distance correlation generated in the effective action i.e. a kinetic term generated from fermion loops that has a low or zero mass.  Such terms can only arise for fluctuations possessing a gauge invariance that through Ward Identites protects against masses of order of the lattice scale. It is certainly true that if we start with two bare fields, one sharing a gauge invariance with the fermion field and one not, the gauge invariant field is the only one that can have low mass and be seen at low energy.  We extrapolate from that clearcut case to assert that the soup that is $H_I$ will select out at low energy a ``fine tuned'' subset of linear combinations of the $H_I$ as an anomaly free set of propagating gauge fields.  

The first consequence is to ensure the metastability criterion requiring that only local fluctuations propagate - since the fermions are local, so will be the  gauge fields since only local gauge invariances can be supported with a local fermion and all the vast number of non-local fluctuations are not observed at low energy.   Therefore the effective low energy interaction among the species, if there is any, can only be an anomaly-free local gauge theory. The only possibilities are gauge invariances that respect the global invariances of the background configuration, namely Poincare invariance and $U(8)_L\otimes U(8)_R$. 

The general possibility of propagating gauge fields being emergent at low energy from a fundamental {\it random dynamics} with many nonlocal degrees of freedom and inhereting locality from local fermion loops has been proposed previously by Nielsen and Rugh \cite{Nielsen1992}.  

Scalar fields linking right and left handed species are probably ruled out, as there is no gauge invariance that can prevent a lattice-scale scalar mass, the well-known problem for fundamental Higgs in non-supersymmetric models. Consequently any non-zero light fermion and gauge field masses  need to be generated by some form of dynamical spontaneous symmetry breaking. 

In the next section a specific dynamical mechanism that may operate to produce fermion and W masses will be discussed. First, it is shown that $H_I$ does contain enough degrees of freedom to produce at least the Standard Model interactions.  It is not immediately obvious that this is so, despite its general random form, since the species are packed together non-trivially into the single spinor through doubling.  

\subsection{Local Bare $U(8)_L \otimes U(8)_R$ symmetry}

First it is shown that the configurations required to turn the global $U(8)_L\otimes U(8)_R$ symmetry of the noninteracting action into a bare local one are present in $H_I$. This group is anomalous and must therefore not be an effective theory at any scale, at least with only the 16 fermions in the $8_L$ and $8_R$ representations. We expect the anomalous currents in $U(8)_L\otimes U(8)_R$ to generate masses of the order of the lattice scale in the corresponding induced propagators and decouple, leaving an anomaly-free subgroup as the effective gauge theory. We note that the  non-anomalous Pati-Salam Model and the Standard Model gauge group are indeed subgroups.  

The task of establishing that these degrees of freedom are present is done if $S_I=\bar{\psi}H_I \psi$ contains terms of the form $\bar{\psi}_x^{d_L}\bar{\sigma}^\mu G^{d d^\prime}_{\mu x}\psi^{d^\prime_L}_x$ and the conjugate expression $\bar{\psi}_x^{d^\prime_L}\bar{\sigma}^\mu G^{d d^\prime \star}_{\mu x}\psi^{d_L}_x$ where $G^{d d^\prime}_{\mu x}$ is slowly varying, for all $d,d^\prime$, and similarly for the right handed equivalent.

First introduce the projection operator 

\begin{equation}
P_{xy}=\prod_a \delta_{x,y+\hat{a}}+2\delta_{xy}+\delta_{y,x+\hat{a}}.
\end{equation}

This has the property that it sends any rapidly varying local function to zero - i.e $\sum_x \bar{\phi_x}(-1)^{x_r}P_{xy} = \sum_y P_{xy}(-1)^{y_r}\phi_y=0$ - and it has no effect on slowly varying functions. This can be used to describe fluctuations that connect two species and no others. Consider $H^{dd^\prime}=A^{dd^\prime} + A^{dd^\prime \dagger}$ where

\begin{equation}
A^{dd^\prime}_{xy}= \bar{f}_d^\dagger (-1)^{d \cdot x} \bar{\sigma}^\mu G^{d d^\prime}_{\mu x} P_{xy} (-1)^{d^\prime \cdot y}f_{d^\prime}^\dagger
\end{equation}

From the properties of $P$, $\bar{\psi}A^{d d^\prime}\psi$ links only $\bar{\psi}^d$ and $\psi^{d^\prime}$, and using $ f^\dagger f = \bar{f}\bar{f}^\dagger=1$ we find (dropping the L label, which is assumed)

\begin{equation}
\bar{\psi}A^{d d^\prime}\psi= \int d^4x \bar{\psi}^d_x \bar{\sigma}^\mu G^{d d^\prime}_{\mu x} \psi^{d^\prime}_x
\end{equation}

and so the first part of the task is done. 

For the conjugate expression, 

\begin{equation}
(A^{dd^\prime\dagger})_{xy}= f_{d^\prime} (-1)^{d^\prime \cdot x} \bar{\sigma}^\mu G^{d d^\prime \star}_{\mu x} S_{xy} (-1)^{d \cdot y}\bar{f}_{d}
\end{equation}

it is clear that $A^{d d^\prime \dagger} $ links only $\bar{\psi}^{d \prime}$ and $\psi^d$. When $d_0=d_0^\prime$, the spinor factors $\bar{f}_{d^\prime} f_{d^\prime}$ and $\bar{f}_{d} f_{d}$ are both either +1 or -1 and so multiply to 1. Accordingly, when $d_0=d_0^\prime$ all the 2 by 2 field matrices in species space linking $\bar{\psi}_d$ and $\bar{\psi}_{d^\prime}$ with $\psi_d$ and $\psi_{d \prime}$ are Hermitian i.e. $1 G_\mu + \tau^iG^i_\mu$ with $G,G^i$ all real, forming the desired U(2) gauge invariance. However when $d_0 \neq  d_0^\prime$, the spinor factors multiply to -1 and so in species space the field matrices are not Hermitian, with $G^1$ and $G^2$ pure imaginary and the gauge invariance, although still present, is not unitary, being of the form $\delta\psi = i\alpha^i\tau^i  \psi$, $\delta\bar{\psi} = -i \alpha^i\tau^i$, with $\alpha^1$ and $\alpha^2$ pure imaginary.  This is a consequence of $\bar{f} \neq f^\dagger$, necessary to remove the negative sign in front of $\partial_0$ in the kinetic operators for the time doublers to obtain the canonical form $\bar{\sigma}^\mu \partial_\mu$. 

Nonetheless it appears that the effective action will be just as if $G^1$ and $G^2$ were real.  The effect of the pure imaginary form is to multiply the $G^1$ and $G^2$  vertices by $i$, which is compensated by a negative sign in the effective gauge propagators that will be generated by fermion loops in the effective action. A simple example of this equivalence can be seen with the ordinary effective action of QED, where substituting $A_\mu \rightarrow iA_\mu$ yields the same Green's functions between the fermions, and the gauge invariance on the fermion field becomes $\psi \rightarrow e^\alpha \psi$ and $\bar{\psi} \rightarrow \bar{\psi}e^{-\alpha}$, for $\alpha$ real.

The same arguments apply to the right sector, and so the configuration $\sum_{d\geq d^\prime}H^{d d^\prime}$  contains the degrees of freedom to support an effective $U(8)_L \otimes U(8)_R$  gauge invariance, and therefore also contains the gauge fields to support the Standard Model gauge group.

\subsection{The Standard Model and Symmetries}

It does not follow from the above analysis that there must be an effective non-anomalous gauge symmetry at some scale, since there is a great surfeit of non-propagating degrees of freedom in $H_I$ that will affect the upper limit of loop integrals, indeed the distillation into separate species itself dissolves at the upper limit, and the possibilty that there are no effective interactions at all has not been ruled out. Nonetheless the existence of the necessary degrees of freedom means that the Standard Model, which is an anomaly free subgroup, is one candidate.

There are some correspondences with symmetries of the species suggesting that the Standard Model subgroup is particularly suited to the left handed lattice Weyl operator, compared with alternative anomaly free subgroups such as Pati-Salam. First, it seems reasonable that there should be a chiral asymmetry in the effective action, with the mass of $W_R$ greater than $W_L$, considering that $H_B$ is fundamentally left handed and the left-handed configurations in $H_I$ will typically be more symmetrical. As an example of a highly symmetrical operator that is chirally selective, the projection operator $P$ selects out only $\nu_L$:  i.e. at low energy $\bar{\psi}P\psi=\bar{\nu}_L\nu_L$.

Table \ref{tab:table1} shows an assignment of quarks and leptons that reflect further correspondences.

\begin{table}[htb]
\caption{\label{tab:table1}One possible set of quark and lepton assignments}
\begin{ruledtabular}
\begin{tabular}{cccccccccc}
 &$d_0$&$d_1$&$d_2$&$d_3$&&$d_0$&$d_1$&$d_2$&$d_3$\\
\hline
$\nu_L$&0&0&0&0&$\nu_R$&1&0&0&0\\
$e_L$&1&1&1&1&$e_R$&0&1&1&1\\
$u_{Lr}$&0&1&1&0&$u_{Rr}$&1&1&1&0\\
$u_{Lb}$&0&1&0&1&$u_{Rb}$&1&1&0&1\\
$u_{Lg}$&0&0&1&1&$u_{Rg}$&1&0&1&1\\
$d_{Lr}$&1&0&0&1&$d_{Rr}$&0&0&0&1\\
$d_{Lb}$&1&0&1&0&$d_{Rb}$&0&0&1&0\\
$d_{Lg}$&1&1&0&0&$d_{Rg}$&0&1&0&0\\

\end{tabular}
\end{ruledtabular}
\end{table}

Full reversal of doubling status in all dimensions is an especially simple transformation, and with the above assignments reversing L(R) doubling status separately corresponds to change of weak L(R) charge. For example, $u_{Lb}$ (d=0101) reverses to $d_{Lb}$ (d=1010).

Further, the discrete reflection and permutation symmetries of the Weyl lattice are cubic, not hypercubic due to the difference between $\sigma^0$ and $\sigma^i$ that gives the Lorentz $SO(3,1)$ invariance of the stationary states. This means that species interactions relating to discrete space dimension symmetries might be expected to survive preferentially. Change of SU(3) color in Table \ref{tab:table1} can be seen to be equivalent to permuting doubling status amongst the space directions only. For example, changing red to blue corresponds in all quark fields to swapping the doubling status in the $2$ and $3$ directions. The electrons and neutrinos, having the same doubling status in all space directions, are singlets under this operation as required.

\subsection{Effective Gauge Propagators and Mass Generation}

What has so far been shown is that the simplest potentially stable configuration of $H$ is the Hermitian Weyl Minkowski space lattice,  that the low energy limit has the same fermion spectrum as the extended Standard Model, that the fluctuations of $H_I$ containenough degrees of freedom for the effective gauge group to be a subgroup of $U(8)_L\otimes U(8)_R$, and that the Standard Model symmetries seem to fit neatly with lattice and species symmetries. 

The low energy effective theory will contain the only interactions that can survive at long distance by generating massless or near-massless propagators, such as the anomaly free set of local gauge invariant interactions discussed above.  

Finding the stationary point of the effective action is equivalent to solving the Schwinger-Dyson equations for the dressed propagators and vertices. 
The dressed quark and lepton propagators in general can acquire mass and  gauge propagators can be generated due to fermion loops. 

\begin{figure}
\includegraphics{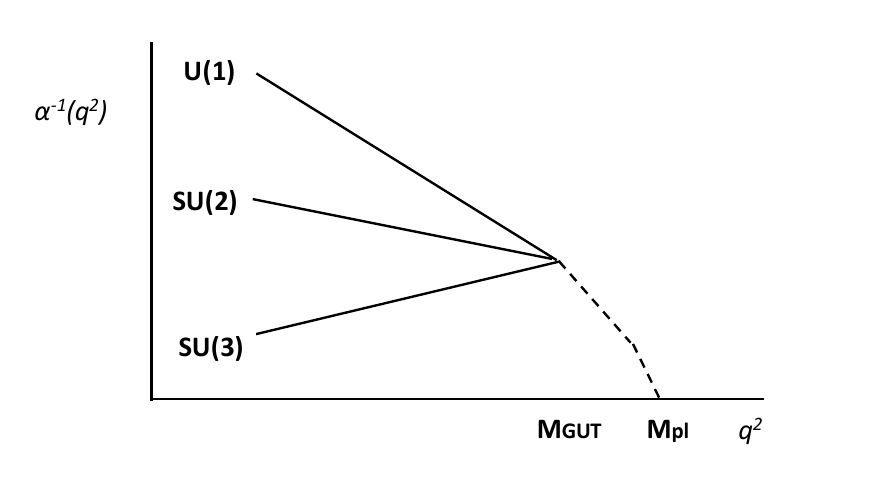}
\caption{\label{fig:running} Emergence of RG running couplings from the strongly coupled theory}
\end{figure}

In a lattice theory with bare gauge kinetic term $\beta F^2$ the full dressed gauge propagator $D$ can be expressed in terms of the full pure gauge field propagator $D_g$ (which goes like $1/\beta$) and the full fermion polarisation tensor $\Pi$ as  

\begin{equation}
D = D_g+ D_g\Pi D_g + D_g\Pi D_g\Pi D_g + \cdots
\end{equation}
summing to $D^{-1}=(1-\Pi D_g) D_g^{-1}$.  In the bare strong coupling limit $\beta=0$, all the gauge propagation is generated by the fermions and the full inverse propagator is

\begin{equation}\label{DFPi}
D^{-1} = -\Pi 
\end{equation}
which is a non-closed form integral equation for $D$ since $\Pi$ contains $D$ in the full vertex. At the Planck lattice scale, scale invariance and therefore the renormalization group is a bad approximation and we expect masses and couplings to be dependent on the high momentum cutoff defined by the Planck scale. At the one-loop level the effective coupling strength near the Planck scale cutoff $\Lambda$ will have the form  
\begin{equation}\label{alpha}
\alpha^{-1}(q^2)= \frac{N_G}{3\pi}\log(\Lambda^2/q^2)
\end{equation}
where $N_G$ is the number of generations, although this could be substantially modified in the full propagator by the many other non-gauge invariant and nonlocal contributions in $H_I$. At longer distance scales where the discreteness of the lattice is not important and all the nonlocal and other non-gauge invariant field contributions have fallen away, the renormalization group analysis will become valid and the full general non-Abelian character including triple gluon vertices will appear in the running coupling from the full vertex in $\Pi$. This scheme may be summarised in Figure \ref{fig:running} for the case that the Standard model emerges from near an intermediate GUT scale, although alternatively the inverse couplings could meet at the strong coupling lattice cutoff $M_{pl}$ .

In QED, strong bare couplings have been studied in many approximations to the Schwinger-Dyson equations and in Euclidean lattice simulations,  where it has been found that when the bare coupling exceeds a critical value a fermion mass is generated, small compared to the UV cutoff when near the critical point\cite{Bloch2002,Atkinson1994,Curtis1990,Kondo1992}.   The bare strong coupling limit $\beta=1/e_0^2=0$ in QED  has been studied in Euclidean lattices where it has been found that fermion mass is generated as long as the number of fermion flavours is less than an apparently critical number of about 30\cite{Dagotto1989}, although another approach suggests an asymptotic behaviour upper limit\cite{Azcoiti1993}. As in (\ref{alpha}) above, when the number of flavours goes up the screening increases, and also the dynamically generated fermion mass will go down. 

Although the details of the fermion loop integrals near the lattice scale will differ from strong coupling QED for our model, a qualitatively similar mechanism to strong QED could operate in our fermionic strong coupling model to generate the current quark and lepton masses. From this chiral symmetry breaking, W masses and composite Higgs-like resonances would result. Gribov proposed such a scheme\cite{Gribov1994}, where the current masses were assumed to be generated from the strong $U(1)_Y$ dynamics at the Landau pole and he showed how the mass of the W-boson and the mass of effective Higgs resonances are a function largely of the heaviest (top) quark mass. Here, all interactions are strong at the Planck scale and so the Landau pole is shifted to the Planck scale.  

\subsection{Gravity} 

Gravity requires identification of the configurations to support a vierbein field and possibly an independent connection field\cite{Percacci2009}, and generation of curvature terms in the effective action from fermion loops as in the gauge field case. Clearly the vierbein will be on the linking entries such as $\delta_{x,y+\hat{\mu}}$ that define the flat background space, where the fluctuations of $H_I$ provide the degrees of freedom to support an effective vierbein field $e_a^\mu$ around the background value $\delta_a^\mu$. The connection field ${A_\mu}^a_b$ may emerge in a more complex combination. 

If the observed low-energy local invariance is restricted to volume-preserving coordinate transformations, as might arise from the unimodular nature of the Lorentz invariance of $H_B$ and/or the fundamental length\cite{Calmet2007} provided by the lattice, unimodular gravity would result, which does not respond to cosmological constant terms and has been proposed as solution of the first cosmological constant problem\cite{Buchmuller1988a,Buchmuller1988b,Weinberg1989,Fiol2008}. 

The manner in which the graviton propagator is proposed to emerge from the bare strong coupling theory of fermions is analogous to Sakharov's idea of induced gravity from one loop dominance, which has recently been shown to produce the right gravitational coupling strength for a fundamental Planck length cutoff when the fermion loops of the 3 generation Standard Model are assumed to dominate\cite{Broda2009}. The approach taken here could provide the theoretical justification for such a calculation.

\section{Generations from Higher Dimensions}

\subsection{Non-quadratic distances}

Higher dimensional lattices can be the source of additional generations. The saturation stability requirement leads to higher dimensions having non quadratic distances. For example, increasing the number of internal components from two to three leads to a saturated matrix of the form $\lambda^\mu_{\alpha\beta}\Delta_{\mu xy}$  where $\lambda^0_{\alpha\beta}=\delta_{\alpha\beta}$ and the other $\lambda^i$ are the eight SU(3) generators.  Whereas the  fermions in Minkowski space move on light cones $p_\mu p_\nu \eta^{\mu\nu}=0$ corresponding to the zero modes of $H$, the zero modes of this nine dimensional space given by $\det(\lambda^\mu p_\mu)=0$ obey a cubic polynomial equation. 

\subsection{Compactified Six component spinor lattice}

When the number of internal components is even, interesting possibilities arise with compactification.  Consider for example the saturated lattice with 6 components and consequently 36 dimensions, which we write in terms of a direct product of 2 by 2 and 3 by 3 spinor matrices as follows:

\begin{equation}
H_{x\alpha i;y\beta j}= i\Delta_{\mu a,{xy}} \lambda^a_{ij}\sigma^\mu_{\alpha\beta}
\end{equation}

Consider that our four dimensions are the four $a=0$ directions, and label the other 4 times 8 directions with $\mu A$, where $A$ ranges from 1 to 8, so that 

\begin{equation}\label{hmulti}
H_{x\alpha i;y\beta j}= i\Delta_{\mu 0,{xy}} \delta_{ij}\sigma^\mu_{\alpha\beta} +i\Delta_{\mu A,{xy}} \lambda^A_{ij}\sigma^\mu_{\alpha\beta}
\end{equation}

Note that time will always be identified with the direction corresponding to the identity matrix in spinor space.

If all the other 32 $\mu A$ directions  have relatively few (or very few) points and are closed then they would have a radius $R$ smaller than measured by current energies. Since the submatrices in the compact directions are relatively small (maybe as small as 3 by 3), it is not the eigenvalue repulsion for the small matrices in isolation that will figure in the stabilising forces but larger crooked submatrices previously discussed, between sets of points travelling down a non-compact direction, taking a turn into the compact direction for a few steps and then turning back into a non-compact direction. Such submatrices will tend to stabilise the points in the compact directions to have same nearest-neighbour links as those of the larger directions, and so the stable operator $\Delta_{x^{A\mu} y^{A\mu}}$ along a compact axis will be simply the closed difference operator. 

The momenta $p_{\mu A}$ (eigenvalues of the $\Delta_{\mu A,{xy}}$) will then be spaced in units of order $1/R$ and at low energies there will not be enough energy to excite transitions between momenta in the compact directions, meaning the partition function will be four dimensional with 3 generations from the first term in (\ref{hmulti}) above. 

Recalling that the Standard Model species arise due to there being two zero modes in each of the large dimensions, there is the question of whether the compact directions produce any more doubling.  Each compact direction that has doubling will double the number of generations seen in Minkowski space up to a maximum of $2^{32}$= 4 billion times!  For an odd number of points the closed difference operator has one zero eigenvalue, and therefore no doubling. For an even number greater than two there are two zero eigenvalues and therefore doubling. However the eigenvalues generally pair for the even sizes and this may destabilize them compared to odd sizes.  

Clearly lattices with 4,8,10,12 or higher-spinors can also be considered.

The additional generations from compact direction doubling can have different effective interactions from the three $a=0$ generations, depending on the form of $H_I$ in the compact direction that survives in the low energy limit.  A form without dependence on the compact direction will lead to the full set of interactions, and a form that has operators like $P$ in the compact directions will decouple or alter the interactions of some or all of the species. Such behaviour might be needed to produce a viable dark matter candidate, since regular standard model Dirac neutrinos appear to be disfavoured by the data\cite{Belanger2008}, and could also generate brane states with only gravity in the extra dimensions, or mirror dark matter\cite{Foot2008}. We also note that additional anomaly-compensating species could allow the high energy gauge group to be expanded to $SU(8)_L\otimes SU(8)_R$  which been reported as a  candidate GUT that can account for the low energy coupling strengths without resorting to supersymmetry\cite{Perez1999}. 

Derivation of the correct shape of the propagating interactions in extra dimensions from the random dynamics is needed to decide amongst the possibiities.

\section{Possible Underlying Principles}
 
\subsection{Infinite Dimensional Conformal Invariance}

If we accept that the observed low energy physics could emerge from the general linear matrix model, why would the fundmental theory be this matrix model? Matrix models rip apart the rules of a local dimensioned spacetime manifold and allow each of the $N$ points to dynamically link with all others. There is clearly an analogy between the general linear transformation and a coordinate transformation in an $N$ (assumed infinite) dimensional complex space, and between the Hermitian matrix and a metric or vielbein over the complex space. 

In Einstein gravity the ordered continuous local spacetime is parametrized by four coordinates $x^\mu$ and the general coordinate invariance $x\rightarrow x^\prime(x)$ confines the spacetime dynamics to the metric $g_{\mu\nu}(x)$ which is needed to have an invariant spacetime displacement $ds^2=g_{\mu\nu}(x)dx^\mu dx^\nu$.  

Consider the same process applied to a pregeometrical theory of $N$ discrete points where each point $a$ is assigned a separate complex number $z^a$ and demand invariance under general analytical transformations $z^a \rightarrow z^{\prime a}(z)$. This yields the transformation of the displacement $dz^{\prime a}= (\partial z^{\prime a}/\partial z^b) dz^b \equiv \Lambda^{-1}(z)^a_b dz^b$ and its conjugate, an invariant ``displacement'' $d\bar{z}^T Hdz$  in the $N$ dimensional complex space is provided by a (non-positive definite) metric Hermitian matrix $H(z,\bar{z})$ transforming as $H \rightarrow \Lambda^\dagger H \Lambda$, and and invariant `volume' \ is $\det H D\bar{z}Dz$.

Generalizing the simple invariant matrix integral $\int\hat{D}H$ of H-theory to the simple conformally invariant functional integral $\int\hat{\cal D}H$ over the metric matrix $H(z,\bar{z})$  doesn't add any new physics, as the $z$ and $\bar{z}$ dependence is a dummy label that can be ignored in the absence of a non-trivial integrand, taking us back to the simple matrix model. Nonetheless it does provide a ``derivation'' of $H$ and its transformation properties as those of a metric required to deliver the conformal invariance. There may also be more complicated invariant integrals involving integrands with actions and curvature terms that should be investigated.

 It is worth noting that string theory manages to deduce local gauge particles and gravitons by demanding conformal invariance over a world sheet embedded in local spacetime parametrized by a single complex number $z$, and in a sense one could interpret this proposal as a generalization of string theory, but the interpretation is quite different on its face and the integrands are trivial. 

\subsection{Bare Particulars}

The arguments of the previous section embed a fundamental `gauge' principle into a pregeometrical  quantum space spanned by discrete entities, half of which end up as points of a lattice spinor spacetime, and the propagating gauge and gravitational fields in spacetime are the remnants of this principle localised by matrix statistics.

The gauge principle is more like a deductive process - we apply numbers to span a configuration space and then deduce the ``cleaning'' forms such as invariant integrals over gauge fields or metrics or matrices because we want the numbers (or parts of them) not to matter.  Why then is this process of application and cleaning fundamental to physics?

As an attempt at an answer, imagine that the entities are no more than a discrete plurality of independent  entities that are maximally simple in having no properties at all. These are not like featureless points represented in a space, that actually have distinguishing spatial coordinates and relations, or like featureless points on a line,  that actually have sequential order and all the arithmetic that follows, but {\it bare} entities that have {\it no intrinsic or relational properties}.  

Bare entities suffer from a representational complexity.   The complexity comes about because conceptual spaces that we are accustomed to think of things as being `in' introduce something from the outside - the coordinate - which is laden with properties.

However representation must involve associating {\it something} with the entity and expressing a form with the something. If we just decided that since bare things are featureless we won't associate anything with them, we would have represented {\it nothing}, not a set of featureless {\it entities}. Associating a set of numbers $\{z\}$ to the set of entities inherently introduces properties, the algebraic relations (e.g. 1+2=3 or $i^2=-1$), so static numbers do not represent bareness. Nonetheless the representational tool has to be algorithmically generated, i.e. some sort of number, and the only way of writing down a form with numbers that has no algebraic relations is to sum over all number values in such a way so that at each point in $z^a$-space, $z^a$ doesn't matter. Hence the need for also associating the metric matrix with pairs $ab$ having its transformation law, and for an invariant sum $\int\hat{\cal D}H$ over the metric that allows the transformation to be realised as a change of variable inside the sum. 

Adopting the continuous complex numbers for $z^a$ and conformal invariance for the symmetry makes some sense as the representational choice because, unlike the integers or reals, the complex numbers are algebraically complete, able to represent all (commutative, associative and distributive) algebraic relations, written as conformal equations $F(z)=0$. One may question whether the complex numbers contain enough algebraic relations to apply and then clean off - what about noncommutative and nonassociative algebraic relations of the quaternions and octonions, or the sedenions? And should we stop at cleaning the entities $a$ represented by $z^a$- what about thinking of the entity pairs $ab$ as another composite entity and cleaning the matrix $H_{ab}$ with another matrix $H_{ab;cd}$? 

There are many uncertainties but the key suggestion to provide a path further forward is that the fundamental entities are very bare, maybe utterly bare, and they defy the arbitrary properties of numbers. 

There is a similarity or perhaps identity of the bare entity concept with the Bare Particulars long considered by some metaphysical philosophers to be required to represent the fundamentally `thingy' essence of entities that make them something other than abstract. The Bare Particular associated with an entity is the entity stripped bare of all the properties it instantiates\cite{Stanford2009}. Sider\cite{Sider2006} has drawn an analogy between entities associated with relational but not intrinsic properties and points of classical spacetime, and a further analogy between what he calls utterly Bare Particulars (entities associated with neither relational nor intrinsic properties) and the singletons of Lewis's set theory.  So perhaps there might be a connection betwen physics and the foundations of set theory, which itself is a foundation of arithmetic. We have already begun to suspect that the foundations of physics and the foundations of mathematics are the same\cite{Tegmark1}.

\section{Conclusion}

The fundamental physical theory ought to provide an explanation for {\it everything}, including Minkowski spacetime, spin, fermions with Pauli statistics, gauge invariance, Standard Model generations, chirality, gravity, mass and even quantum field theory itself. Some of these features can be anthropically selected but the theory has to tell us which ones.

H-theory potentially provides the explanations of these features in sometimes unexpected ways and arguably points much more strongly to our familiar laws than string theory, although by conformal invariance it might be related. It might seem unsatisfactory that such a numerically pathological expression as $\int \hat{D}H$ could be proposed as a fundamental expression, but pathology may be an inevitable price to pay to model bare entities using numbers. Alternatively there may be a better formulation that also generates a finite matrix model of the general form $\int DH \det H$ on which all the predictions depend.  

In the scenario painted here, physics is modelled by the general linear matrix model because the fundamental entities are Bare Particulars, requiring representation with complex numbers in an expression having infinite dimensional conformal invariance. Spacetime is not a consequence of the quantization of curvature that leads to disordered spacetime foams, but of the enormously strong statistics of the nonlocal Hermitian matrices that lead to ordered robust local lattices of first order difference operators. The Minkowski Weyl 3+1 dimensional spinor lattice has a central role as the only potentially stable multidimensional configuration with quadratic distances. Time is the difference operator associated with ${\bold 1}$ in spinor space. Fermionic Grassmann matter fields are not a consequence of supersymmetry but are a remnant determinant from the underlying pregeometrical symmetry. The 16 Weyl spinors of the Standard model generations are unavoidable and intimately linked with the lattice spacetime via species doubling. Generations and possibly dark matter are from compactified higher dimensions. Local gauge and gravitational fields are surviving low energy anomaly-free effective gauge theories from a soup of general random fluctuations strongly interacting with the local fermions.

There are many directions for future work to explore the several unproven conjectures and possibilities presented here, including: 

\begin{itemize}
\item rigorous derivation of the reduced system; 
\item the stability criteria for lattices; 
\item the random dynamics mechanism ; 
\item emergence of the Standard Model gauge group; 
\item fermion and W mass patterns; 
\item vierbein propagator and the cosmological constant; 
\item cosmological consequences of the locally flat robust background;
\item effective propagators in the extra dimensions; and
\item representational consequences of the bare entity hypothesis.  
\end{itemize}

Some computer simulations might be tractable although representing the full non-local set of interactions will severely restrict the achievable lattice size. Clearly a major positive outcome would be to confirm that the Standard Model gauge group arises as the long distance effective interaction between the 16 species.

\begin{acknowledgments}
The author acknowledges useful discussions with Robert Foot regarding anomalies and Pamela Tate and Anita Avramides regarding the philosophical work on Bare Particulars.

\end{acknowledgments}

\newcommand{\hepth}[1]{\href{http://xxx.lanl.gov/abs/hep-th/#1}{\tt hep-th/#1}}
\newcommand{\hepph}[1]{\href{http://xxx.lanl.gov/abs/hep-ph/#1}{\tt hep-ph/#1}}
\newcommand{\arXivid}[1]{\href{http://arxiv.org/abs/#1}{\tt arXiv:#1}}
\newcommand\npb[3]{{Nucl.\ Phys.\ }{\bf B#1}, #3 (#2)}
\newcommand\npbps[3]{{Nucl.\ Phys.\ Proc.\ Supp.}{\bf B#1}, #3 (#2)}
\newcommand\annp[3]{{Ann.\ Phys.\ (NY)\ }{\bf #1}, #3 (#2)}
\newcommand\plb[3]{{Phys.\ Lett.\ }{\bf B#1}, #3 (#2)} 
\newcommand\pprd[3]{{Phys.\ Rev.\ }{\bf D#1}, #3 (#2)} 
\newcommand\pprl[3]{{Phys.\ Rev.\ Lett.}{\bf #1}, #3 (#2)} 
\newcommand\jcap[3]{{JCAP}{\bf #1}, #3 (#2)} 
\newcommand\jhep[3]{{JHEP}{\bf #1}, #3 (#2)} 
\newcommand\mpla[3]{{Mod. Phys. Lett.}{\bf A#1}, #3 (#2)} 
\newcommand\rrmp[3]{{Rev. Mod. Phys.}{\bf #1}, #3 (#2)} 
\newcommand\jmp[3]{{J. Math. Phys.}{\bf #1}, #3 (#2)}

\end{document}